# Improvement of ISOM by using filter

Imen Chaabouni [1], Wiem Fourati[2] and Med Salim Bouhlel[3]

[1] Sciences and Technologies of Image and Telecommunications High Institute of Biotechnology
Sfax, Tunisia
chaabouni.imen@gmail.com

[2] Sciences and Technologies of Image and Telecommunications High Institute of Biotechnology
Sfax, Tunisia
wiem_fourati@yahoo.fr

[3] Sciences and Technologies of Image and Telecommunications High Institute of Biotechnology
Sfax, Tunisia
medsalim.bouhlel@enis.rnu.tn

**Abstract**

Image compression helps in storing the transmitted data in proficient way by decreasing its redundancy. This technique helps in transferring more digital or multimedia data over internet as it increases the storage space. It is important to maintain the image quality even if it is compressed to certain extent. Depend upon this the image compression is classified into two categories : lossy and lossless image compression. There are many lossy digital image compression techniques exists. Among this Incremental Self Organizing Map is a familiar one. The good pictures quality can be retrieved if image denoising technique is used for compression and also provides better compression ratio. Image denoising is an important pre-processing step for many image analysis and computer vision system. It refers to the task of recovering a good estimate of the true image from a degraded observation without altering and changing useful structure in the image such as discontinuities and edges. Many approaches have been proposed to remove the noise effectively while preserving the original image details and features as much as possible. This paper proposes a technique for image compression using Incremental Self Organizing Map (ISOM) with Discret Wavelet Transform (DWT) by applying filtering techniques which play a crucial role in enhancing the quality of a reconstructed image. The experimental result shows that the proposed technique obtained better compression ratio value.

***Keywords:*** *Image compression, Incremental Self Organizing Map, Discret Wavelet Transform, Filter, Denoising.*

## 1. Introduction

Image denoising has been one of the most important and widely studied problems in image processing and computer vision. The need to have a very good image quality is increasingly required with the advent of the new technologies in a various areas such as multimedia, medical image analysis, aerospace, video systems and others. Indeed, the acquired image is often marred by noise which may have a multiple origins such as: thermal fluctuations; quantify effects and properties of communication channels. It affects the perceptual quality of the image, decreasing not only the appreciation of the image but also the performance of the task for which the image has been intended such as image compression which is a result of applying data compression to the digital image. The main objective of image compression is to decrease the redundancy of the image data which helps in increasing the capacity of storage and efficient transmission. Image compression aids in decreasing the size in bytes of a digital image without degrading the quality of the image to an undesirable level. Different compression approaches have been studied in the literature, for example, the Incremental Self Organizing Maps (ISOM)[15]. There are many types of self-organizing networks applicable to a wide area of problems. One of the most basic schemes is competitive learning. The Kohonen network can be seen as an extension to the competitive learning network [9]. Self Organizing Maps are a kind of artificial neural networks which inspire from the learning neural networks. This kind of neural network allows projecting an entry space on a one or two dimensional map called topological map. It's composed of two layers, an entry vector, and a map where all elements are of the same dimension as for the entry[13].

## 2. Image Compression with combined ISOM and DWT2

### 2.1. Data compression using SOM

One important feature of SOM is the possibility of achieving high compression ratio with relatively small block size. Another important advantage of SOM image compression is its fast decompression phase by rebuilding the compressed image. SOM is basically a clustering method, grouping similar vectors (blocks) into one class [6]. Our basic approach to image compression consists of several key steps.

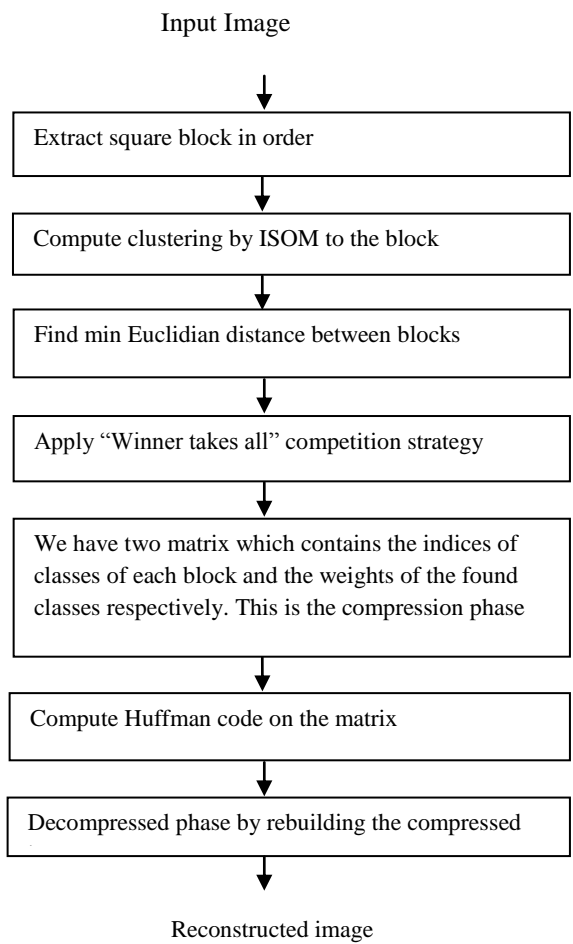

Fig.1: Algorithm of compression

### 2.2. The Discret Wavelet Transform

The DWT provides sufficient information for the analysis and synthesis of a signal, but is advantageously, much more efficient. Discrete Wavelet analysis is computed using the concept of filter banks[7], [10], [11].

### 2.3. Compression techniques

There are many different forms of data compression [14]. This investigation will concentrate on transform coding and then more specifically on Wavelet Transforms.

Image data can be represented by coefficients of discrete image transforms. Coefficients that make only small contributions to the information contents can be omitted. However wavelets transform is applied to entire images, rather than subimages, so it produces no blocking artefacts. This is a major advantage of wavelet compression over other transform compression methods.

This explains that the wavelet analysis does not actually compress a signal, it simply provides information about the signal which allows the data to be compressed by ISOM which use standard entropy coding techniques, such as Huffman coding. Huffman coding is good to use with a signal processed by wavelet analysis, because it relies on the fact that the data values are small and in particular zero, to compress data. It works by giving large numbers more bits and small numbers fewer bits. Long strings of zeros can be encoded very efficiently using this scheme. Therefore an actual percentage compression value can only be stated in conjunction with an entropy coding technique.

### 2.4. Implementation of 2D DWT with ISOM

The basic steps used in the ISOM compression were:

(1) Calculate the DWT of the image

(2) Compute the ISOM technique on the first bloc of decomposition wavelet (the approximations)

(3) Apply the inverse discrete wavelet transform (IDWT) for the decompressed image

## 3- Image pretreatment with filters

Data sets collected by image sensors are generally contaminated by noise. Imperfect instruments, problems with the data acquisition process, and interfering natural phenomena can all degrade the data of interest. Furthermore, noise can be introduced by transmission errors and compression. Thus, denoising[12] is often a necessary and the first step to be taken before the image data is analyzed. It is necessary to apply an efficient denoising technique to compensate for such data corruption.

## 3.1 Spatial Filtering

A traditional way to remove noise from image data is to employ spatial filters. Spatial filters can be further classified into non-linear and linear filters.

### a. Non-Linear Filters

With non-linear filters, the noise is removed without any attempts to explicitly identify it. Spatial filters employ a low pass filtering on groups of pixels with the assumption that the noise occupies the higher region of frequency spectrum. Generally spatial filters remove noise to a reasonable extent but at the cost of blurring images which in turn makes the edges in pictures invisible. In recent years, a variety of nonlinear median type filters such as weighted median [4], rank conditioned rank selection [2], and relaxed median [5] have been developed to overcome this drawback.

### b. Linear Filters

A mean filter is the optimal linear filter for Gaussian noise in the sense of mean square error. Linear filters too tend to blur sharp edges, destroy lines and other fine image details, and perform poorly in the presence of signal-dependent noise. The wiener filtering [1] method requires the information about the spectra of the noise and the original signal and it works well only if the underlying signal is smooth. Wiener method implements spatial smoothing and its model complexity control correspond to choosing the window size. To overcome the weakness of the Wiener filtering, Donoho and Johnstone proposed the wavelet based denoising scheme in [3], [8].

## 4- Experimental results

In order to evaluate the performance of the proposed approach of image compression using modified ISOM with DWT2 and filters, two standard images are considered. The work is implemented using MATLAB. Cameraman.tif in figure 2 and Peppers.bmp in figure 3 are the two standard images used to explore the performance of the proposed approach of image compression. The experiments are carried out with the block size 8*8.

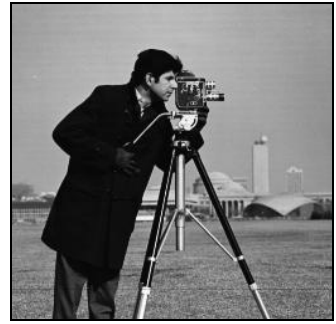

Fig.2 : Cameraman.tif

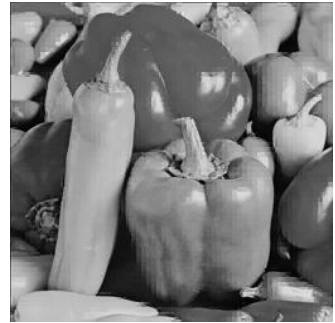

Fig.3 : Peppers.bmp

The evaluation of the proposed approach was performed using the following measures,

$$\text{MSE} = \|y - x\|^2 = \frac{1}{N*N} \sum_{i=1}^{N} = \sum_{j=1}^{N} (x_{i,j} - y_{i,j})^2 \quad (1)$$

$$\tau = (1 - \frac{T_c}{T_o}) \times 100 \quad (2)$$

Where $x_{i,j}$ and $y_{i,j}$ are the pixel intensities for the original and the reconstructed image, $T_c$ and $T_o$ are the size of compressed and original file and $N*N$ is the size of the image.

These two factors will decide about the Mean Square Error (MSE) and the ratio of compression ($\tau$) for the digital image. The MSE is the cumulative squared error between the compressed and the original image, whereas $\tau$ is the compression ratio between the compressed and the original image.

The experimental results that evaluate the performance of the proposed approach by comparing it with filters are tabulated. Table 1 and table 2 shows the experimental results applied for Cameraman.tif and Peppers.bmp.

From table 1 and table 2, it can be observed that when applying the four filters respectively (median filter, gaussian filter, mean filter and adaptative filter), we can improve the compression ratio but the mean square error is degraded little. That is compression ratio for Cameraman.tif without filter and using filter respectively (median filter, gaussian filter, mean filter and adaptative filter) is 85,54% , 86,91% , 86,32% , 86,32% and 88,08% whereas the mean square error for this image without filter and using filter respectively is 54,66 , 58,95 , 60,25 , 100,9 and 118,4. With this analysis it can be said that the compression ratio obtained after using filter will be better than this without using filter, especially with using adaptative filter. Unfortunately, the considered noise in the reconstructed image is highly when using filter.

For the second image Peppers.bmp, the compression ratio is increasing when using filter. Without using filter we have 77,92 % and by using filter respectively (median filter, gaussian filter, mean filter and adaptative filter) we have 78,51% , 78,12% , 78,41% and 79,1%. This clearly indicates that the compression ratio obtained after using filter will be better than the compression ratio obtained without using filter. But the mean square error is little degraded when using filter from 60,8 without using filter to 66,7 , 61,09 , 62,28 and 65,11 by using respectively (median filter, gaussian filter, mean filter and adaptative filter) filter.

Table 1: Experimental results for Cameraman.tif

| Image compression without filter | Image compression with median filter | Image compression with gaussian filter | Image compression with mean filter | Image compression with adaptative filter |
|---|---|---|---|---|
| $\tau = 85,54\ \%$<br>MSE = 54,66 | $\tau = 86,91\ \%$<br>MSE = 58,95 | $\tau = 86,32\ \%$<br>MSE = 60,25 | $\tau = 86,32\ \%$<br>MSE = 100,9 | $\tau = 88,08\ \%$<br>MSE = 118,4 |
| 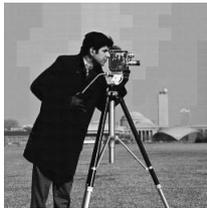 | 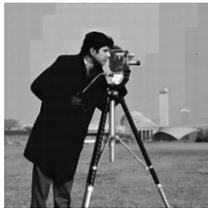 | 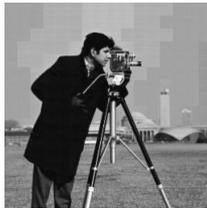 | 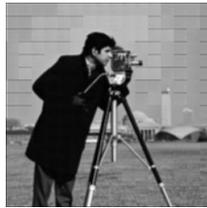 | 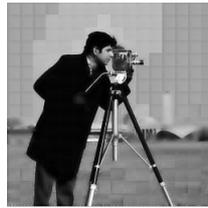 |

Table 2: Experimental results for Peppers.bmp

| Image compression without filter | Image compression with median filter | Image compression with gaussian filter | Image compression with mean filter | Image compression with adaptative filter |
|---|---|---|---|---|
| $\tau = 77,92\ \%$<br>MSE = 60,8 | $\tau = 78,51\ \%$<br>MSE = 66,7 | $\tau = 78,12\ \%$<br>MSE = 61,09 | $\tau = 78,41\ \%$<br>MSE = 62,28 | $\tau = 79,1\ \%$<br>MSE = 65,11 |
| 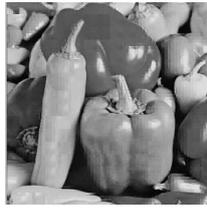 | 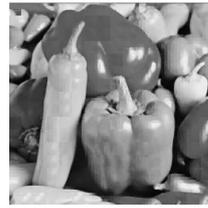 | 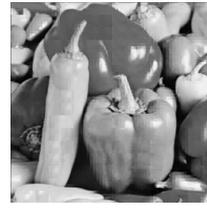 | 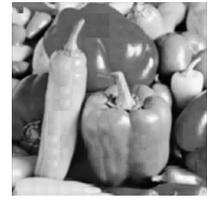 | 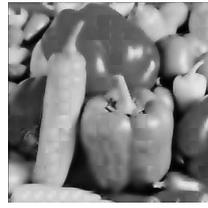 |

## 5- Conclusion

Incremental Self Organizing Map is a popular learning based method and has been widely applied for image compression. This proposed paper introduced standard filter respectively (median filter, gaussian filter, mean filter and adaptative filter) for ISOM with DWT algorithm for image compression.

We verify the compromise that exists between the compression ratio and the quality of the rebuilt image. Indeed, we apply the different filter respectively (median filter, gaussian filter, mean filter and adaptative filter), when we use ISOM to the compression technique. We note that the incremental self organizing map with DWT transform compression method provide us better results when using filter than without using. When we apply filter to ISOM with DWT compression technique the compression ratio and the mean square error is increasing.

## References
[1] A.K.Jain,Fundamentals of digital image processing. Prentice-Hall,1989
[2] R. C. Hardie and K. E. Barner, "Rank conditioned rank selection filters for signal restoration," IEEE Trans. Image Processing, vol. 3, pp.192–206, Mar. 1994.
[3] David L. Donoho and Iain M. Johnstone,"Ideal spatial adaption via wavelet shrinkage", Biometrika, vol.81, pp 425-455, September 1994
[4] R. Yang, L. Yin, M. Gabbouj, J. Astola, and Y. Neuvo, "Optimal weighted median filters under structural constraints," IEEE Trans. Signal Processing, vol. 43, pp. 591–604, Mar. 1995.
[5] A. Ben Hamza, P. Luque, J. Martinez, and R. Roman, "Removing noise and preserving details with relaxed median filters," J. Math. Imag. Vision, vol. 11, no. 2, pp. 161–177, Oct. 1999.

[6] Teuvo Kohonen, Self Organizing Maps. Springer, Berlin, Heidelberg, Third Extended Edition (2001).
[7] Maarten Jansen, Noise Reduction by Wavelet Thresholding, Springer –Verlag New York Inc.- 2001.
[8] David L. Donoho and Iain M. Johnstone., "Adapting to unknown smoothness via wavelet shrinkage", Journal of the American Statistical Association, vol.90, no432, pp.1200-1224, December 1995. National Laboratory, July 27, 2001.
[9] Luger F., (2002), Artificial Intelligence: Structures and Strategies for complex problem solving, 4th edition, Addison-Wesley, 856p.
[10] Lakhwinder Kaur, Savita Gupta, and R. C. Chauhan, Image Denoising using Wavelet Thresholding, Third Conference on Computer Vision, Graphics and Image Processing, India Dec 16-18, 2002.
[11] Savita Gupta and Lakhwinder kaur, Wavelet Based Image Compression using Daubechies Filters, In proc. 8th National conference on communications, I.I.T. Bombay, NCC-2002.
[12] Y. Yang, Image denoising using wavelet thresholding techniques 2005.
[13] Zumray D., (2006), A unified framework for image compression and segmentation, Expert Systems with Applications..
[14] Durai, S-A, An improved image compression approach with SOFM Network using Cumulative Distribution Function, Août 2007, Advanced Computing and Communications, p304-307.
[15] Chaabouni I., (2010), Image compression with Self Organizing Maps, IJCIIS, January 2010Issue, p64-71.

**Imen Chaabouni** received the D.E.A degree from University of Sfax, Tunisia, in 2007 in Automatics Industrial Computing and she work in thesis to University Sfax, Tunisia, from 2010, in Image Processing. She is currently a Ph. D Student of compute engineering at ENIS.

Ph. D Student Chaabouni has been working on Image processing, since 2005. His current interests include the image compression and pretreatment. (Research Unit on Sciences of Electronics, Technologies of Information and Telecommunications (SETIT), University of Sfax, Sfax Engineering School (ENIS), BP W, 3038 Sfax, Tunisia.

Phone: +261 25 98 41 82, E-mail: (chaabouni.imen@gmail.com)

**Wiem Fourati** received the D.E.A degree from University of Sfax, Tunisia, in 2005 and the Ph.D from university of Sfax, Tunisia in 2007.

**Mohamed Salim Bouhlel** is actually the Head of Biomedical imagery Department the Higher Institute of Biotechnology Sfax (ISBS).He has received in 1999 the golden medal with the special mention of jury in the first International Meeting of Invention, Innovation and Technology (Dubai). He was the Vice President of the Tunisian Association of the Specialists in Electronics. He is an associate professor at the Department of Image and Information Technology in the Higher National School of Telecommunication ENST-Bretagne (France).